\documentclass[prl,twocolumn,showpacs,debugon,nofootinbib]{revtex4}
\usepackage{epsfig}
\newcommand{\bk}{{\bf k}}
\newcommand{\be}{\begin{equation}}
\newcommand{\ee}{\end{equation}}
\renewcommand{\r}{{\bf r}}
\newcommand{\s}{{\bf s}}

\newcommand{\q}{{\bf q}}

\newcommand{\R}{{\bf R}}

\begin{document}


\title{Interference of diffusing photons and level crossing spectroscopy}
\author{E. Akkermans and O. Assaf}
\affiliation{Department of Physics, Technion Israel Institute of Technology,
  32000 Haifa, Israel}

\begin{abstract}
We show that a new interference effect appears in the intensity fluctuations of photons multiply scattered by
an atomic gas of large optical depth $b$. This interference occurs only
for scattering atoms that are Zeeman degenerate and it leads to a deviation from  the Rayleigh law. The fluctuations measured by their variance, display a resonance
peak as a function of an applied magnetic field.  The resonance width is
proportional to  the small factor $1/b$. We derive closed analytic expressions for all these physical quantities which are directly accessible experimentally.

\end{abstract}


\date{\today}

\maketitle


We consider in this letter coherent multiple scattering of photons propagating in cold atomic gases. For Zeeman degenerate atoms, we show that there is a new interference effect which affects the fluctuations of intensity (speckle pattern) of  photons transmitted through a gas of large optical depth, but leaves unchanged the average transmitted intensity. This interference shows up in a significant deviation from the Rayleigh law which states that for classical scatterers  without internal structure, the variance $\overline{\delta {\cal
T}^2}=\overline{{\cal T}^2}-\overline{{\cal T}}^2$ of the transmission coefficient $\cal T$ is simply related to its average value $\overline{\cal T}$ by $\overline{\delta
{\cal T}^2}=\overline{{\cal T}}^2$. The averaging over configurations denoted by $\overline{\cdot \cdot \cdot}$ will be defined later. The interference involves the ground state Zeeman sublevels of the atoms. 
The dependence of this interference on magnetic quantum numbers suggests  that it is sensitive to an applied magnetic
field $H$. Our purpose in this letter is to show that indeed the
amplified variance presents a resonance as a function of $H$ around a crossing point,
and the width of this resonance is \be \Delta H \simeq a { \hbar \Gamma \over g \mu_0}
{l \over L} \label{eq1} \ee where $l$ is the photon elastic mean
free path through the atomic gas confined into a slab of width
$L$, $g$ is the Land\'e factor and $\mu_0$ the Bohr
magneton. Here, $a$ is a constant of order unity to be
determined later, that depends on the details of the atomic
structure. All these features are obtained in the limit of diffusing photons, {\it i.e.}, in a regime where $L\gg l$. Thus the narrowing of the resonance is in
principle not limited, which might prove useful in level-crossing spectroscopic measurements. We shall see that the effect we present here shares some kind of analogy with the well-known Franken or Hanle effects \cite{arimondo,franken}. Nevertheless, we emphasize that both the underlying physical mechanisms and the quantities that are being measured are very different from these two effects \cite{brossel,series}. 

\begin{figure}[ht]
\centerline{ \epsfxsize 7cm \epsffile{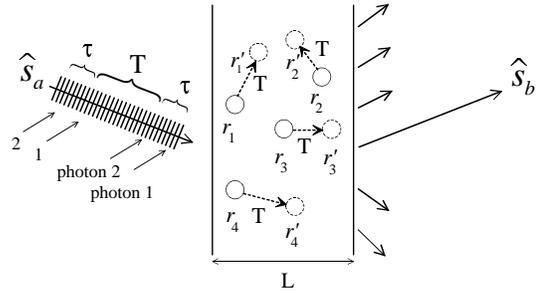} } \caption{\em
Experimental setup. A CW laser beam is incident onto atoms confined in a slab of width $L$. A detector placed along the direction ${\hat \s}_b$, records the intensity. During the time interval $\tau$, assumed to be short, the scatterers stay at rest. However, the two photons 1 and 2 experience different atomic internal configurations due to all other photons between them. The measurement is repeated after a time $T \gg \tau$. }
\label{Setup}
\end{figure}

We consider the setup of Fig.\ref{Setup}. A photon of polarization
$\hat {\bf \varepsilon}_a$ is incident along a direction ${\hat
\s}_a$ onto the atomic gas. It is detected in transmission
with polarization $\hat {\bf \varepsilon}_b$, along ${\hat \s}_b$
after being multiply scattered. A time $\tau$ later, a second
identical photon is detected. We assume that $\tau$ is short enough
so that the atoms stay at rest between the two events. The same
measurement is repeated after a time $T \gg \tau$, during which
the scatterers move. The averaging over spatial disorder results
from this motion. The transmitted intensity
${\cal T}$  is proportional to the probability of a photon incoming along ${\hat \s}_a$, to emerge
along ${\hat \s}_b$.
Atoms are modeled as degenerate two-level systems. We denote by
$| m_{g} \rangle$ the ground state Zeeman sublevels with total angular momentum $j_g$, $|m_{e}
\rangle$ are the excited states sublevels with total angular momentum $j_e$, and $m$ is the  projection on a quantization axis. 

The average transmission coefficient $\overline{\cal T}$ is
obtained by squaring the sum of the scattering amplitudes, $A_n^{\{R,m\}}$,
corresponding to a given configuration $\{R,m\}$. Here $\{R\}$
accounts for the spatial positions of all atoms, $\{m\}$
is a notation for their internal Zeeman states both before and
after scattering and the index $n$ denotes one possible multiple
scattering path. Then, 
\be \overline{\mathcal{T}}= \overline{
\Bigl|\sum_{n}A_n^{\{R,m\}}\Bigr|^2} = \sum_{n
n'}\overline{A_n^{\{R,m\}}A_{n'}^{\{R,m\}*}}  \label{transav}\ee
where $\overline{\cdot \cdot \cdot}$ denotes a configuration
average over both $\{R\}$ and $\{m\}$. When averaging over
$\{R\}$, all cross terms $n\neq n'$ vanish because of large
fluctuating phase shifts, so that
$\overline{\mathcal{T}} \simeq \sum_{n} \overline{
|A_n^{\{m\}}|^2}$. This expression is the leading approximation in
the weak disorder limit $k_0 l \gg 1$, where $k_0$ is the photon
wave number  \cite{am}. The two photons detected at $t=0$ and $t=\tau$ are separated by many undetected photons (see Fig.\ref{Setup}) which may change the internal states of atoms. Therefore, if $\{m\}$
and $\{m'\}$ refer to the atomic internal configuration seen by the two detected photons, then we can assume that
there is no correlation between $\{m\}$ and $\{m'\}$.

Similarly, the correlation function of the transmission coefficients $\cal T$ and ${\cal T}'$ of  the two detected photons is
\be
\overline{{\cal T}{\cal T}'} = 
\sum_{ijkl} \overline{A_i^{\{R,m\}}A_j^{\{R,m\}*}A_k^{\{R,m'\}}A_l^{\{R,m'\}*}}
\label{transfluc}
\ee
As before, the averaging over $\{R\}$ leaves only pairs of
amplitudes having exactly opposite phase shifts. Thus, to leading
order in weak disorder, the only non vanishing contributions
involve two possible pairings of amplitudes, either $i=j,k=l$,
which gives ${\overline{\cal T}} \, \overline{{\cal T}'}$, or $i=l,j=k$, so that defining ${\cal C}^2 =\overline{{\cal T}{\cal T}'}-{\overline{\cal T}} \, \overline{{\cal T}'}$, we obtain
 \be
{\cal C}^2 =
\sum_{i j} \overline{
A_i^{\{m\}}A_i^{\{m'\}*} A_j^{\{m'\}}A_j^{\{m\}*}}
\ee
This correlation function appears as products of two amplitudes, that correspond to different internal configurations $\{m\}$ and $\{m'\}$, but to
the same scattering path $i$ (or $j$). Most of multiple scattering paths $i$ and $j$ do not share common  scatterers so that  we can average $A_i^{\{m\}}A_i^{\{m'\}*}$ and
$A_j^{\{m'\}}A_j^{\{m\}*}$ separately, since these averages are
taken upon different atoms, and finally,
\be
{{\cal C}^2} =  \Bigl| \sum_i \overline{
A_i^{\{m\}}A_i^{\{m'\}*}} \Bigr|^{2} \, .
\label{correl1}
\ee
Since generally $\{m\}\neq\{m'\}$, the
interference occurs between distinct Zeeman sublevels of the {\it ground state},
unlike the Franken or Hanle effects, where the interference involves 
distinct {\it excited} sublevels. This constitutes a new kind of
interference, which originates from the fact that it is the
correlation function rather than the average intensity, that is considered.

In the theory of multiple scattering it is helpful to use a
continuous description \cite{am}. In this
framework, one defines two {\it Diffuson}
functions $\mathcal{D}^{(i,c)}$ by \cite{rk2}
\be
\overline{\cal T} = \int d \r d \r'
 {\cal D}^{(i)} (\r, \r') \, \  \ \mbox{and} \, \ \  {\cal C} = \Bigl|
\int d \r d \r'  {\cal D}^{(c)} (\r, \r') \Bigr|
\label{intensitedalbedo12}
\ee
The two functions ${\cal D}^{(i,c)}$ are obtained from an
iteration equation (also called ladder diagram) whose structure is based on two elementary vertices  ${\cal V}^{(i,c)}$, that
describe the microscopic details of the scattering process. The
iteration of the elementary vertices is written symbolically as \be {\cal D} =  {\cal V} + {\cal V}
{\cal W} {\cal V} + \cdots  = {\cal V} + {\cal D} {\cal W}  {\cal
V} \label{iterationD} \ee where ${\cal D}$,${\cal V}$ stand for ${\cal D}^{(i,c)}$,${\cal V}^{(i,c)}$. 
${\cal V} $ accounts for a
single scattering and $ {\cal D} {\cal W} {\cal V}$ represents its
iteration. The quantity $\cal W$ describes the propagation of the
photon intensity between successive scattering events and it will
be described later on.

Generally, the elementary vertex is obtained by coupling two scattering
amplitudes. It is given by
\be
 {\cal V} =
  \sum_{m_i m_e m_e '}{\langle
m_2|V(\hat {\bf \varepsilon}_1,\hat {\bf \varepsilon}_2)|m_1\rangle \langle m_4|V(\hat {\bf \varepsilon}_3,\hat {\bf
\varepsilon}_4)| m_3\rangle^\ast \over (
\omega-\omega_{m_1m_e}+i{\Gamma \over 2})(\omega-\omega_{m_3m_e'}-i{\Gamma \over 2})} \label{vertexeq} \ee where the operator $ V(\hat {\bf
\varepsilon}',\hat {\bf \varepsilon}) =  \hat {\bf
\varepsilon}'^{*} \cdot{\bf d}|m_e \rangle\langle m_e |
{\bf d}  \cdot \hat {\bf \varepsilon} $ results from the dipolar
interaction energy $ -{\bf d}.{\bf E}$ between atoms and photons. 
$\bf d$ and $\bf E$ are respectively the atomic dipole and  electric
field operators. The states $|m_i\rangle$ are Zeeman sublevels of
the atomic ground state, and $|m_e\rangle ,|m_e'\rangle$ are
those of the excited state. We have defined the
energy difference $\hbar \omega_{ij} = E_j - E_i$ and the photon frequency $\omega$. We assume that the ground state Zeeman
sublevels are equiprobable so that the corresponding density
matrix reduces to the factor $1/(2j_g+1)$.

The elementary vertex ${\cal V} ^{(i)}$, that corresponds to the
average intensity, is obtained by setting $m_1
= m_3$, $m_2 = m_4$, $\hat {\bf \varepsilon}_1 = \hat {\bf
\varepsilon}_3$ and $\hat {\bf \varepsilon}_2 = \hat {\bf
\varepsilon}_4$ in (\ref{vertexeq}). Up to a proportionality
factor, $ {\cal V}^{(i)} $ is nothing but the differential cross
section for this scattering process.  Assuming a broad line
excitation \cite{rkbroad}, we average ${\cal V} ^{(i)}$ over
$\omega$, leading to \cite{franken}
 \be
 {\cal V} ^{(i)} =
 \sum_{m_1 m_2} \sum_{m_e m_e'} {B_{12} (m_e) B_{12}^* (m'_e) \over  i \, \omega_{m_e m_e'} +   \Gamma} \label{ivertex}
 \ee
where $B_{12} (m_e) = \langle m_2| \hat {\bf
\varepsilon}_2^{*} \cdot{\bf d}|m_e \rangle\langle m_e | {\bf d}
\cdot \hat {\bf \varepsilon}_1|m_1\rangle$. A magnetic field removes  the level degeneracy and leads to
a Zeeman splitting, so that two kinds of terms appear in (\ref{ivertex}) depending on
whether $m_e = m'_e$ or $m_e \neq m'_e$. Terms for which $m_e =
m'_e$, are independent of magnetic field and give the
incoherent scattering cross section.  The terms $m_e \neq m'_e$
depend on magnetic field and describe interferences between
two distinct scattering amplitudes.

\begin{figure}[ht]
\centerline{ \epsfxsize 5cm \epsffile{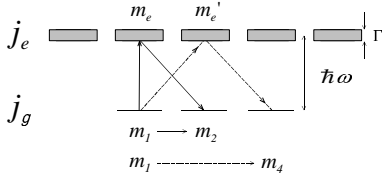} } \caption{\em
Example of a term which, for the transition $j_g =1 \rightarrow j_e =2$, contributes to $\mathcal{V}^{(c)}$, but not to $\mathcal{V}^{(i)}$. The solid and dashed lines refer respectively to the two quantum amplitudes that appear in (\ref{vertexeq}). For large enough magnetic field, this contribution vanishes since Zeeman splitting takes it far from resonance. }
\label{Vert}
\end{figure}

For the vertex ${\cal V}^{(c)}$, each one of the two
coupled scattering amplitudes in (\ref{vertexeq}) might belong to a
distinct atomic configuration (see (\ref{correl1})), meaning that we must consider
distinct couples of initial $(|m_1\rangle , | m_3 \rangle)$
and final $(|m_2\rangle , | m_4 \rangle)$ atomic states, as
well as two initial $(\hat {\bf \varepsilon}_1, \hat {\bf
\varepsilon}_3)$ and final $(\hat {\bf \varepsilon}_2, \hat {\bf
\varepsilon}_4)$ polarization states. Summations
over the quantum numbers $m_i$ result from averaging over initial
atomic states and from non detected final states. The vertex ${\cal V}^{(c)}$ involves more interference terms than those already appearing in ${\cal V}^{(i)}$. A non degenerate ground state leads immediately, using  (\ref{vertexeq}), to ${\cal V}^{(i)} = {\cal V}^{(c)}$ so that we recover the Rayleigh law,  ${{\cal C}^2}= {\overline{\cal T}} \,  \overline{{\cal T}'}$ \cite{am}. Degenerate states produce additional interference terms in ${\cal V}^{(c)}$ so that ${{\cal C}^2} > {\overline{\cal T}} \,  \overline{{\cal T}'}$ (see Fig.\ref{Vert}).  An applied magnetic field removes the degeneracy and therefore affects the interference pattern.

We now calculate the Diffusons $\cal D$ given by
the iteration (\ref{iterationD}). We first decompose the various terms into components in a standard basis, 
 \be {\cal V} = \sum_{\alpha \beta \gamma
\delta} (\hat {\bf \varepsilon}_1)_{-\alpha} (\hat {\bf
\varepsilon}_2)_\gamma ^* (\hat {\bf \varepsilon}_3)_{-\beta} ^*
(\hat {\bf \varepsilon}_4)_\delta \, \ {\cal V}_{\alpha \beta,
\gamma \delta} \, .\label{vcomposantes} \ee Likewise, Eq.(\ref{iterationD}) acquires a tensorial
form,
 \be
 {\cal D}_{\alpha \beta, \gamma \delta} =  {\cal V}_{\alpha \beta, \gamma \delta}  + W \sum_{\mu \nu \rho \sigma}  {\cal D}_{\alpha \beta, \mu \nu} {\cal P}_{\mu \nu , \rho \sigma} {\cal V}_{ \rho \sigma , \gamma \delta} \, \ .
\label{it2} \ee Here $W$
describes the scalar part of $\cal W$ and ${\cal P}_{\alpha \beta, \gamma
\delta} = \langle \left(\delta_{\alpha\gamma} - (-)^{\gamma}{\hat
s}_\alpha {\hat s}_{-\gamma} \right) \left(\delta_{\beta \delta} -
(-)^{\beta}{\hat s}_{-\beta} {\hat s}_\delta \right) \rangle$
accounts for the polarization dependent part. This follows at once by noticing that after being scattered by an atom,
the two outgoing photon amplitudes propagate with a wavevector
${\hat {\s}} = \bk / k_0$, random in direction but identical for
both, and with two different polarization components. Since ${\hat
{\s}}$ is random, the intensity propagation is averaged $\langle
\cdot \cdot \cdot \rangle$ over photon wavevectors direction. The
term $\delta_{\mu \nu} - (-)^{\nu}{\hat s}_\mu {\hat s}_{-\nu}$
expresses transversality. The two tensors ${\cal P}_{\alpha \beta,
\gamma \delta}$ and  ${\cal V}_{\alpha \beta, \gamma \delta}$ can
be written as $9 \times 9$ matrices. The iteration (7) rewrites 
${\cal D}=(1+W{\cal VP}+(W{\cal VP})^2+\cdot\cdot\cdot){\cal V}$. We now use the spectral decomposition theorem to expand ${\cal VP}=\sum_K u_K T^{(K)}$, where $u_K$'s are the
eigenvalues of ${\cal VP}$ and the $T^{(K)}$'s define an orthonormal set of (generally) 9 projectors  \cite{ohadbig}. Then, with the help of (\ref{it2}), we find
 \be
 {\cal D}_{\alpha \beta, \gamma \delta} ^{(i,c)} = \sum_K U_K ^{(i,c)} \left({\cal V}_K ^{(i,c)} \right)_{\alpha \beta, \gamma \delta}
\label{eq13} \ee with $ {\cal V}_K ^{(i,c)} = T^{(K)} {\cal
V}^{(i,c)}$ and \be U_K ^{(i,c)} 
 \simeq  { 8 \pi c \Lambda \over 3 l^2}  {1 /  u_K
^{(i,c)} \over { \gamma_K ^{(i,c)}} + D q^2 } \label{eq14}
\ee where $\Lambda = (2 j_e +1) / 3 (2 j_g +1)$ and $\q$ (with $q=|\q|$) is the Fourier variable of the
difference $\R=\r'-\r$ between the two endpoints of a multiple
scattering sequence. The {\it r.h.s} in (\ref{eq14}) is obtained
by using the diffusion approximation ({\it i.e.} $ql \ll 1$), so
that $W(q) \simeq {3 \over 2 \Lambda} (1 - q^2 l^2 / 3)$,
where $D= c l / 3$ is the photon diffusion coefficient \cite{am}.
We identify the set of characteristic damping rates \be \gamma_K ^{(i,c)} =
{c \over l} \left( {2 \Lambda \over 3 u_K ^{(i,c)}} -1 \right) \, .
 \label{tauk} \ee
The term $\frac{2}{3}\Lambda$ is the total
cross section conveniently normalized. According to the values of $u_K ^{(i,c)}$, we
identify 3 kinds of modes. A positive $\gamma_K
^{(i,c)}$ describes an exponentially damped mode. A
vanishing $\gamma_K ^{(i,c)}$ corresponds to an infinite
range stable mode which ensures energy conservation, and a negative damping rate describes an
amplified mode. The largest eigenvalue of ${\cal V}^{(i)}{\cal P}$
is $u_0^{(i)}=2\Lambda/3$, thus leading to one stable mode. For
degenerate scatterers ($j_g,j_e>0$), and without magnetic field,
$\mathcal{D}^{(c)}$ has one amplified mode, whose
occurrence results from the fact that all terms that contribute to
$\mathcal{V}^{(i)}$, contribute also to $\mathcal{V}^{(c)}$ \cite{ohadbig}.
However, there are interference terms that contribute to
$\mathcal{V}^{(c)}$ only (Fig.\ref{Vert}).
Therefore, the largest eigenvalue of ${\cal V}^{(c)} {\cal P}$
becomes greater than $\frac{2}{3}\Lambda$, thus making the corresponding damping rate negative.

We now rewrite (\ref{eq13})  in real space,  \be {\cal D}^{(i,c)} (\r,\r')
=   \sum_K Y_K ^{(i,c)} \int_0^{\infty} dt \,
\  {\cal D}(\r,\r',t) \, \ e^{- \gamma_K ^{(i,c)} t} \label{it3}
\ee 
where $Y_K ^{(i,c)}$ are two angular functions that depend on the incoming and outgoing polarizations ${\hat {\bf \varepsilon}}_a$ and ${\hat {\bf \varepsilon}}_b$ \cite{ohadbig}. 
The
scalar Diffuson propagator ${\cal D}(\r,\r',t)$ obeys a diffusion
equation whose solution for a slab geometry is well known
\cite{am} and leads for (\ref{intensitedalbedo12}) to \be
{\cal C}  = \sum_{K} Y_K ^{(c)} \,{\sinh^2 (l/L_K ^{(c)}) \over ( l/ L_K ^{(c)})
\sinh (L / L_K ^{(c)})}
\label{corr} \ee
where we have defined $ L_K ^{(c)} = \sqrt{ D /
\gamma_K ^{(c)} } $. The average transmission coefficient $\overline{\cal T}$ is given by the same relation (\ref{corr}) provided we replace $(L_K ^{(c)},Y_K ^{(c)})$ by $(L_K ^{(i)},Y_K ^{(i)})$ defined accordingly. 

The dominant contribution to the average intensity
$\overline{\mathcal{T}}$ is the stable, energy conserving mode $\gamma_0^{(i)}=
0$. The two other modes have
positive damping rates and are negligible compared to this stable mode. They express photon depolarization in multiple
scattering.  
The stable mode leads to $\overline{{{\cal T}}} \propto l/L=1/n\sigma L$. Here $n$
is the density of scatterers and $\sigma$ is the single scattering
total cross section. However, the
total cross section is independent of the magnetic field
$H$ \cite{franken}. This can be understood as follows. Starting from (\ref{ivertex}), the
outgoing polarization dependent part is a sum of terms like $\sum_{\hat{\varepsilon}_2 \perp {\bf k}}\langle
m_2|\hat{\varepsilon}_2^\ast\cdot {\bf d}|m_e\rangle\langle m_e '
|\hat{\varepsilon}_2\cdot {\bf d}|m_2\rangle=\sum_{\bf k
}(\varepsilon_{2\alpha}\varepsilon_{2\beta}+\varepsilon'_{2\alpha}\varepsilon'_{2\beta})\langle
m_2 |d_\alpha |m_e\rangle\langle m_e ' |d_{-\beta}|m_2\rangle$,
where ${\bf k}$ is the outgoing wave vector, and $(\alpha,\beta)$ are the standard components of $\bf d$ that contribute to the transitions. Integrating over ${\bf k}$ imposes $\alpha=-\beta$,
which implies $m_e=m_e '$. Thus the interference terms in (\ref{ivertex}) 
does not contribute to $\sigma$, which is therefore independent  of $H$. As a consequence, $\overline{{\cal T}}$ is also independent of $H$.

\begin{figure}[ht]
\centerline{ \epsfxsize 6cm \epsffile{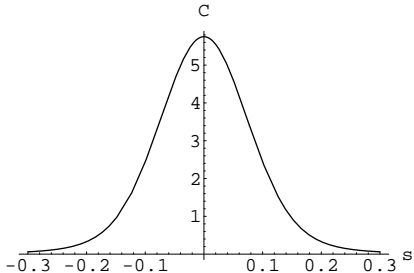} } \caption{\em
Plot of ${\cal C}$ given by (\ref{fluct2}) as a function of the
dimensionless magnetic field $s$. We denote by $\Delta$ its FWHM.
This plot corresponds to the $j_g =1 \rightarrow
j_e =2$ transition and $b =5$. } \label{DeltaT}
\end{figure}

\begin{figure}[ht]
\centerline{ \epsfxsize 6cm \epsffile{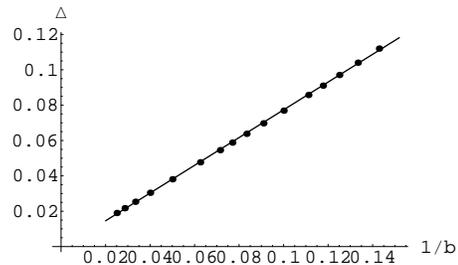} } \caption{\em
FWHM $\Delta$ of ${\cal C}$ plotted versus $1/b$. The points are obtained from (\ref{fluct2}) in which $f_2$ results from a direct numerical calculation of $\gamma_0 ^{(c)} (s) $. The slope of this linear behavior is in good agreement with the predicted expression $2 \sqrt{\ln 2 / f_2} \simeq 0.76$ obtained for the transition $j_g =1 \rightarrow j_e =2$. }
\label{Halfwidth}
\end{figure}

The intensity correlation function $\cal C$ is dominated by the amplified mode driven by the negative damping rate $\gamma_0 ^{(c)}$. The integral in (\ref{it3}) is cutoff by $t_{max} = {\cal L} / c $ where ${\cal L} = c L^2 / D$ is the longest path of a diffusing photon. Eq.(\ref{corr}) thus leads to \cite{ohadbig}
\begin{equation}
{\cal C}=
Y_0^{(c)}
\left({\sin^2({X \over b}) \over {{X}\sin X}}-2\sin^2({\pi
 \over b})\frac{e^{- \pi ^2 + X^2}}{\pi ^2- X^2} \right)
\label{fluct2}
\end{equation}
where $X= {L / L_0 ^{(c)}}$ and $b = L/ l$ is the optical depth. This expression is displayed in Fig.\ref{DeltaT} as a function of the dimensionless magnetic field $s= g \mu_0 H / \hbar \Gamma$.
It retains the shape of a resonance curve whose FWHM is given by (\ref{eq1}). To derive it, we expand ${\cal V}^{(c)} (s) $ to first order in $s$, leading for the amplified mode to $u_0 ^{(c)} (s) \simeq u_0 ^{(c)} (0) - \beta s^2$ where $\beta$ is a constant that depends on the specific scattering atom. According to (\ref{tauk}), the corresponding damping rate becomes,
\be
\gamma_0 ^{(c)} (s) \simeq  \gamma_0 ^{(c)} (0) +  {2 \beta \, c  \, \Lambda \over 3  u_0 ^{(c)2} \, l } s^2
\label{gamcs}
\ee
By rewriting $X = b \sqrt{l^2 \gamma_0 ^{(c)} (s) / D}$, and making use of (\ref{gamcs}), we obtain $X = b \sqrt{|f_0 - f_2 s^2|}$, where the two constants $f_0$ and $f_2$ are given by $f_0 = (2 \Lambda / u_0 ^{(c)}) -3$ and $f_2 = 2 \Lambda \, \beta \, / u_0 ^{(c)2}$. For large enough optical depth $b = L / l$, {\it i.e.}, in the diffusive regime where expression (\ref{fluct2}) applies, the FWHM $\Delta$ behaves linearly with $1/ b$ as shown in Fig.\ref{Halfwidth}. The corresponding slope is easily calculated from (\ref{fluct2}) and restoring units, we obtain for $\Delta H$ the expression (\ref{eq1}) with $a = 2 \sqrt{ \ln 2 / f_2}$. 

To summarize, we have shown that in multiple scattering, interference of diffusing photons scattered by atoms in the presence of a magnetic field near a level crossing (or close to zero field) shows up as a resonance peak in the intensity correlation function ${\cal C}$. Its width, given by Eq.(\ref{eq1}), is inversely proportional to the optical depth $b = L / l$. The diffusive regime corresponds to large values of $b$ (typically $b \simeq 10^2$), so that the sensitivity of the interference to a magnetic field is significantly enhanced. This could be used towards more precise measurements in level crossing spectroscopy, in the limit of dense atomic gases where multiple scattering cannot be neglected anymore.

It is a pleasure to thank C. Cohen-Tannoudji for his interest in this work and his comments. This research is supported in part by the Israel Academy of
Sciences and by the Fund for Promotion of Research at the
Technion.


\begin{thebibliography}{99}
\bibitem{rk2} All  prefactors that depend on the geometry of the sample and on the strength and shape of the photon source have been absorbed in the definitions of ${\cal D}^{(i,c)}$.
\end{thebibliography}
    \end{document}